%% file: main.tex
\documentclass[12pt]{article}
\usepackage[version=3]{mhchem}
\usepackage[left=2cm, right=2cm, top=2cm, bottom=2.0cm]{geometry}
\usepackage{caption}
\usepackage{subcaption}
\usepackage{sectsty}
\usepackage[super,sort&compress,comma]{natbib} 
\usepackage{graphicx} 
\usepackage{hyperref}
\usepackage{epstopdf}

\input{math_commands}

\usepackage{authblk}
\title{Enhanced Deep Potential Model for Fast and Accurate Molecular Dynamics; Application to the Hydrated Electron}
\author[1]{Ruiqi Gao}
\author[2]{Yifan Li}
\author[2]{Roberto Car}
\affil[1]{Department of Electrical and Computer Engineering, Princeton University}
\affil[2]{Department of Chemistry, Princeton University}
\date{}                     

\begin{document}
\maketitle

\begin{abstract}
    In molecular simulations, neural network force fields aim at achieving \emph{ab initio} accuracy with reduced computational cost. 
This work introduces enhancements to the Deep Potential network architecture, integrating a message-passing framework and a new lightweight implementation with various improvements.
Our model achieves accuracy on par with leading machine learning force fields and offers significant speed advantages, making it well-suited for large-scale, accuracy-sensitive systems. 
We also introduce a new iterative model for Wannier center prediction, allowing us to keep track of electron positions in simulations of general insulating systems. 
We apply our model to study the solvated electron in bulk water, an ostensibly simple system that is actually quite challenging to represent with neural networks. 
Our trained model is not only accurate, but can also transfer to larger systems. 
Our simulation confirms the cavity model, where the electron's localized state is observed to be stable. 
Through an extensive run, we accurately determine various structural and dynamical properties of the solvated electron.
\end{abstract}

\section{Introduction}
Molecular dynamics (MD) simulations provide insights for physical and chemical processes at the atomic level and have wide applications. 
To perform a simulation under Newtonian motion for the atoms, one needs to calculate the forces, which, under the Born-Oppenheimer approximation, are many-body functions of the atomic coordinates.
Non-empirical quantum mechanical methods such as Density Functional Theory (DFT) can in principle obtain these forces with good accuracy in many situations, but the high computational cost limits such methods to small systems and short time scales.
There also exist classical force fields that are empirical, simple approximations to the many-body force function, which are many orders of magnitude faster to compute and scale linearly with the system size, but often fall short on the accuracy side.
In recent years, machine learning (ML) force fields have become a promising direction to combine the advantages of both sides. That is, they are trained for \emph{ab initio} level of accuracy, while achieving a linear scaling speed. They can still be somewhat slower than classical force fields,
but they are much more scalable and faster than DFT and have been widely used in large-scale simulations.

There has been a lot of development of ML force fields over the years\cite{behler2007generalized, schutt2017quantum, gilmer2017neural, han2017deep, gasteiger2020directional, 
hu2021forcenet, gasteiger2021gemnet, ying2021transformers, liu2022spherical, batzner20223, liao2022equiformer}. 
With the popularity of accuracy benchmarking, more recent models \cite{gasteiger2021gemnet, batzner20223, liao2022equiformer}
generally follow a trend of increasing accuracy at the cost of increasing model and computational complexity. 
However, in MD simulations, equilibrium and dynamical properties may require timescales of nanoseconds or even microseconds, corresponding to millions to billions of steps for sufficiently large system sizes.
To this end, more lightweight and faster models are required.
Earlier models like the Behler-Parrinello Neural Network (BPNN)\cite{behler2007generalized}
and Deep Potential (DP)\cite{han2017deep, zhang2018deep, zhang2018end, jia2020pushing, lu202186} model are relatively small and fast, but may be inadequate for accuracy-sensitive systems. 

This work is focused on developing a model that runs fast while being accurate enough for MD simulations.
It is based on the DP model and we have made various enhancements to it. 
The most important is the incorporation of a message passing (MP) mechanism, so we call it DP-MP. This enables a richer representation that learns features on top of features, and also effectively increases DP's cutoff radius of the local receptive field.
Using most of the building blocks of the existing DP model, we propagate both scalar and vector features for each atom and retain the model's invariance to translation, rotation, and permutation. 
We also incorporate second-order tensor information in the final features. 
These enhancements are designed to significantly boost the accuracy of DP without incurring much computational cost.

To make it faster and more flexible, we implement the new scheme with JAX\cite{jax2018github}, a Python-based autograd and machine learning framework that is optimized on GPUs.\footnote{Code available at \url{https://github.com/SparkyTruck/deepmd-jax}.}
The MD part can be seamlessly connected with frameworks like JAX-MD\cite{schoenholz2019jax}, enabling an end-to-end GPU workflow in Python.
We perform a simple benchmark on a water system. Combined with the new implementation gains, the new model is around two orders of magnitude faster than other models achieving similar accuracy.

Additionally, in this work, we also present a new method for the prediction of the Wannier centers, i.e., the centers of maximally localized Wannier distributions\cite{marzari2012maximally}. Wannier centers can be seen as representing the centers of the charge associated to individual electrons.
So far, the Wannier centers can be predicted by a similar neural network like the DP model,
which is called the Deep Wannier (DW) model\cite{zhang2020deep}.
But this scheme is limited to systems where the Wannier centers can be uniquely associated with individual atoms,
which precludes modeling electron transfer processes.
In the present approach, we encapsulate a prediction model in an iterative refinement process, and it is called DWIR (Deep Wannier Iterative Refinement).
With DWIR, one can keep track of the electrons in an atomic simulation, even when they are not uniquely associated to individual atoms.

To illustrate the capabilities of our enhanced models, we apply them to the study of \ce{e-}(aq), the solvated electron in bulk water. \ce{e-}(aq) plays an important role in radiation chemistry and biology\cite{herbert2017hydrated}, and despite its apparent simplicity, it had undergone much research effort before the cavity model became well-established: The electron creates a localized quasi-spherical cavity with a shell of surrounding water molecules\cite{herbert2017hydrated, ambrosio2017electronic}.
This system poses considerable challenges for ML models since they only see the atoms and not the excess electron, and the structure is quite complex and sensitive compared to bulk water. 
There has been efforts to learn an ML model of \ce{e-}(aq)\cite{lan2021simulating}, but it remains difficult to obtain a sufficiently accurate and robust model\cite{lan2021simulating, lan2022temperature}. 
Also, there has not been a model that can be transferred to larger systems, which is actually a requirement for many applications and technically possible given the localized nature of the electron.

In this work, we perform a DFT simulation of a periodic box of 128 \ce{H2O} molecules plus one \ce{e-}, and use the DP-MP scheme to successfully learn a model of \ce{e-}(aq). 
We demonstrate its transferability to a larger system of 256 \ce{H2O} molecules and one \ce{e-}. 
In the DFT calculations, we use the hybrid PBE($\alpha$) hole functional with 40\% exact exchange and rVV10 van der Waals correction, which has been proved to give a good description for water and \ce{e-}(aq)\cite{ambrosio2016structural, ambrosio2017electronic}. 
We perform a nanosecond-long DP-MP run to collect sufficiently converged statistics, and learn an additional DWIR model to track the position of \ce{e-}(aq). 
We calculate various structural and dynamical properties including the size, radial distribution functions, and diffusion mechanism.
Our calculation confirms the cavity model and the stability of the localized state. We also identify a form of H-\ce{e-} bond around the electron that is similar to the H-bond in water, and whose forming and breaking gives rise to the rapid diffusion of \ce{e-}(aq).

\section{Methodology}

In this section we describe our new models. In Sec.~\ref{sec:DP} we give a recap on the DP model. We introduce the DP-MP model in Sec.~\ref{sec:DP-MP}, and the Wannier-center model DWIR in Sec.~\ref{sec:DW}.

\subsection{Structure of the DP model}
\label{sec:DP}

Given a system of $N$ atoms with coordinates $\{\mathbf{r}_i\}_{i=1}^{N}$, the DP model represents the potential energy surface (PES) as a sum of atomic contributions, each term depending only on the atom's neighboring environment within a cutoff radius $r_c$:
\[ E (\bfr_1, \cdots, \bfr_N) = \sum_i E_{\omega}(\{\bfr_{ij}\}_{j\in \mathcal{N}_{r_c}(i)}) \]
where $\omega$ represents all the learnable parameters of the model, $\bfr_{ij}$ is the relative displacement between atoms $i$ and $j$, and $\mathcal{N}_{r_c}(i)$ is the set of neighboring atoms $j$ for which $r_{ij} < r_c$. The forces are subsequently derived as the gradient of the energy. Each term $E_{\omega}$ is computed as follows:
\begin{enumerate}
    \item Compute a smooth function $s(r_{ij})$ that approximates $\frac{1}{r_{ij}}$, except that it is modified to become zero when $r_{ij} \geq r_c$.
    \item An embedding neural network $G$ takes each ${s(r_{ij})}$ as input and outputs the feature $G(s(r_{ij}))$, a vector of length $M_1$.
    \item Average over neighboring atoms to obtain a scalar ($T^{(1)}$) and vector ($T^{(3)}$) feature of length $M_1$ for each atom $i$:
    \begin{align*}
         T_i^{(1)} &= \frac{1}{N_{nbr}}\sum_{j \in \mathcal{N}_i} s(r_{ij})G(s(r_{ij}))\\
    T_i^{(3)} &= \frac{1}{N_{nbr}}\sum_{j \in \mathcal{N}_i} s(r_{ij})\hat{\bfr}_{ij}G(s(r_{ij}))
    \end{align*}
    where $N_{nbr}$ is a precomputed constant that stands for the average number of neighbors, and $\hat{\bfr}_{ij}$ is the normalized $\bfr_{ij}$.
    \item Obtain an invariant feature $D_i$ of size $M_1M_2$: This is done by taking a subset of $M_2$ ($< M_1$) ``axis'' features from $T$, and for each $m_1$ in $\{1,\cdots, M_1\}$ and $m_2$ in the subset we compute
    \[ 
    D_{i,(m_1, m_2)} = T_{i,m_1}^{(1)} T_{i,m_2}^{(1)} + \langle T_{i,m_1}^{(3)}, T_{i,m_2}^{(3)}\rangle_3
    \]
    where $\langle\cdot, \cdot\rangle_3$ is the inner product over the spatial dimension.
    \item Apply a fitting neural network $F$ that yield the atomic energy 
    \[ E_i = E_\omega(\{\bfr_{ij}\}_{j\in \mathcal{N}_{r_c}(i)}) = F(D_i). \]
\end{enumerate}

The model designed this way preserves the translational, rotational, and permutational symmetry of the energy function. All trainable parameters lie in the embedding network $G$ and fitting network $F$, which are both multi-layer fully-connected residual networks (ResNets)\cite{he2016deep}.
In practice, depending on the chemical species, an embedding network is trained for each type-pair of $(i,j)$ and a fitting network is trained for each type of $i$. But for simplicity we omit the atomic type in the formulas. 

The calculation of the embedding network in step 2 is performed for pairs of atoms, making it the most time-consuming step. Thankfully, the input $s(r_{ij})$ is one dimensional, so it can be approximated by a piecewise polynomial at inference time, or referred to as \emph{compressed}\cite{lu2022dp}. With DP's simple design as well as compression, it is very fast compared to recent ML force fields.

DP has found many successful applications in systems like water, silicon, metal, metal oxides and so on, and has been applied to many studies including the phase diagram, and processes involved in crystal nucleation, combustion, interfacial systems etc.\cite{bonati2018silicon, gartner2020signatures, niu2020ab, andrade2020free, zeng2020complex, zhang2021phase}. However, being a simple model, the expressive power of DP is somewhat limited. In addition, in more complex systems such as \ce{e-}(aq), the radius of its influence most probably extends beyond the usual cutoff where DP is seen to perform well (like $6$~\AA\ in water). This brings us to the enhanced design, described in the next subsection.

\subsection{Enhanced DP Model with Message Passing}
\label{sec:DP-MP}

\begin{figure}[h]
          \centering
          \includegraphics[clip, trim=16.8cm 3.3cm 25.5cm 3.3cm, width=0.30\textwidth]{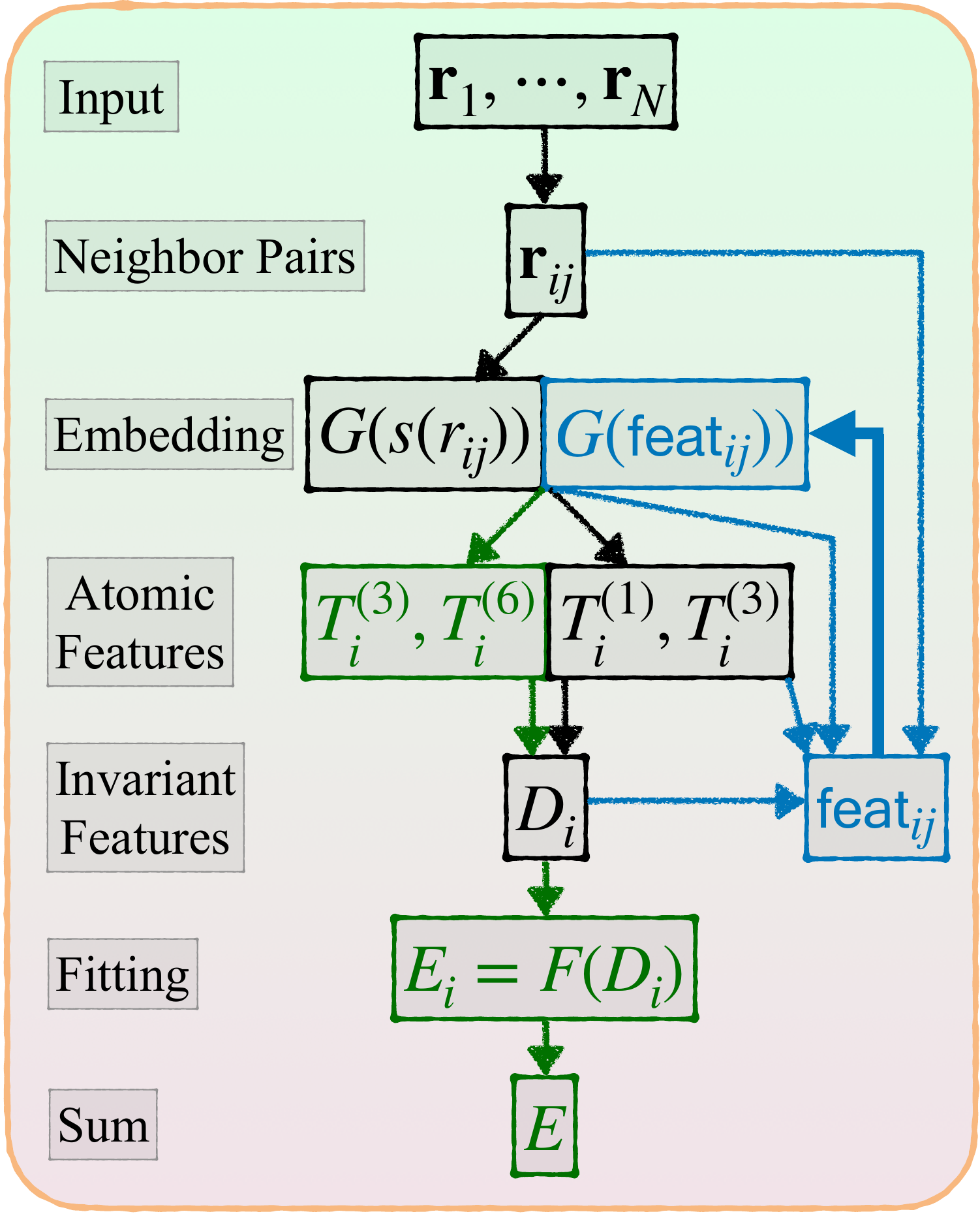}
         \caption{Architecture of DP-MP. The blue parts indicate the message passing steps, and the green parts indicate the final loop.}
         \label{fig:mp}
\end{figure}

\begin{figure}[h]
    \centering
    \begin{subfigure}[b]{0.165\textwidth}
         \includegraphics[clip, trim=24.7cm 5.2cm 24.7cm 5.2cm, width=\textwidth]{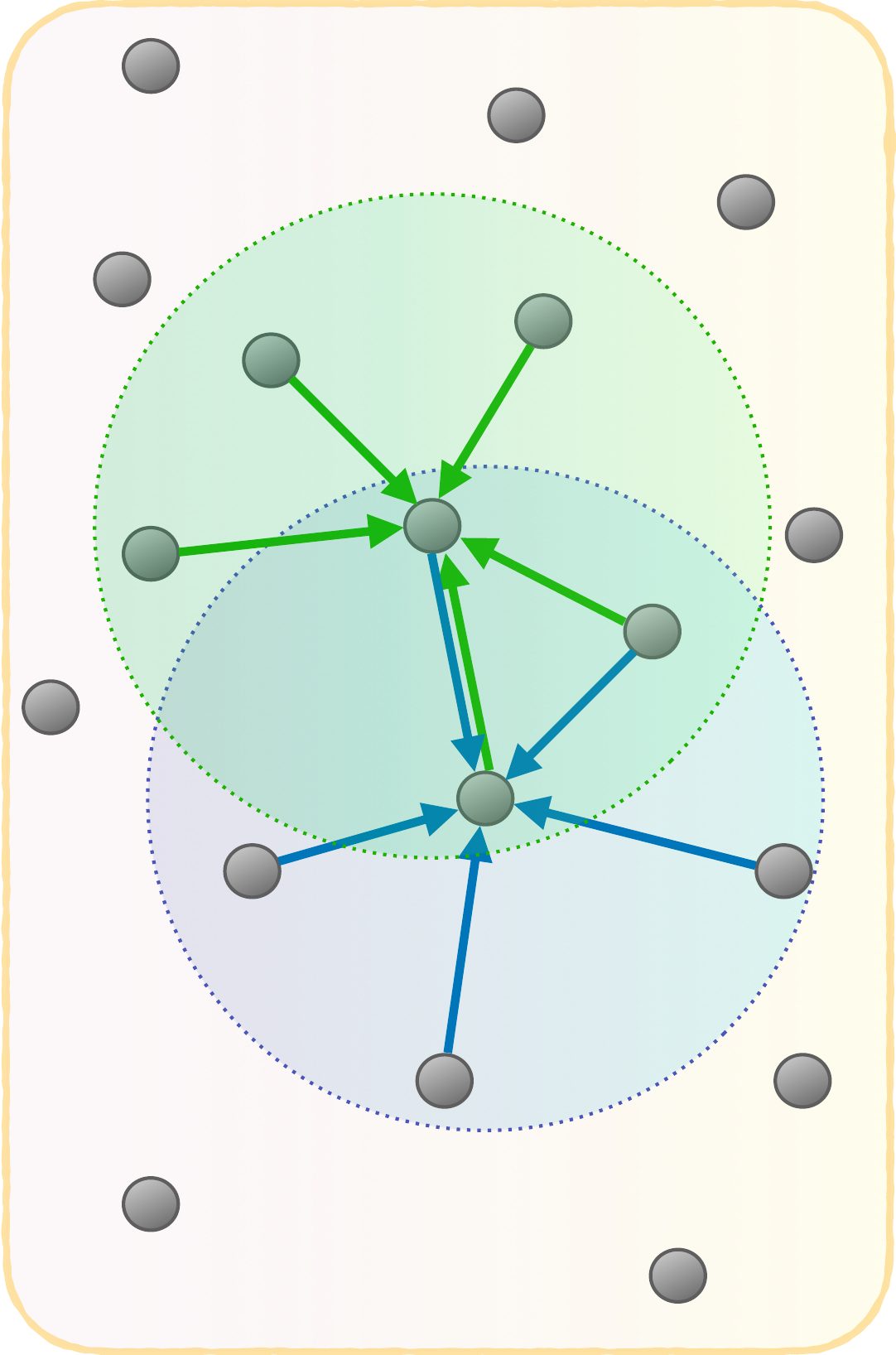}
     \end{subfigure}
     \begin{subfigure}[b]{0.183\textwidth}
        \includegraphics[clip, trim=26.3cm 7.8cm 25cm 7.8cm, width=\textwidth]{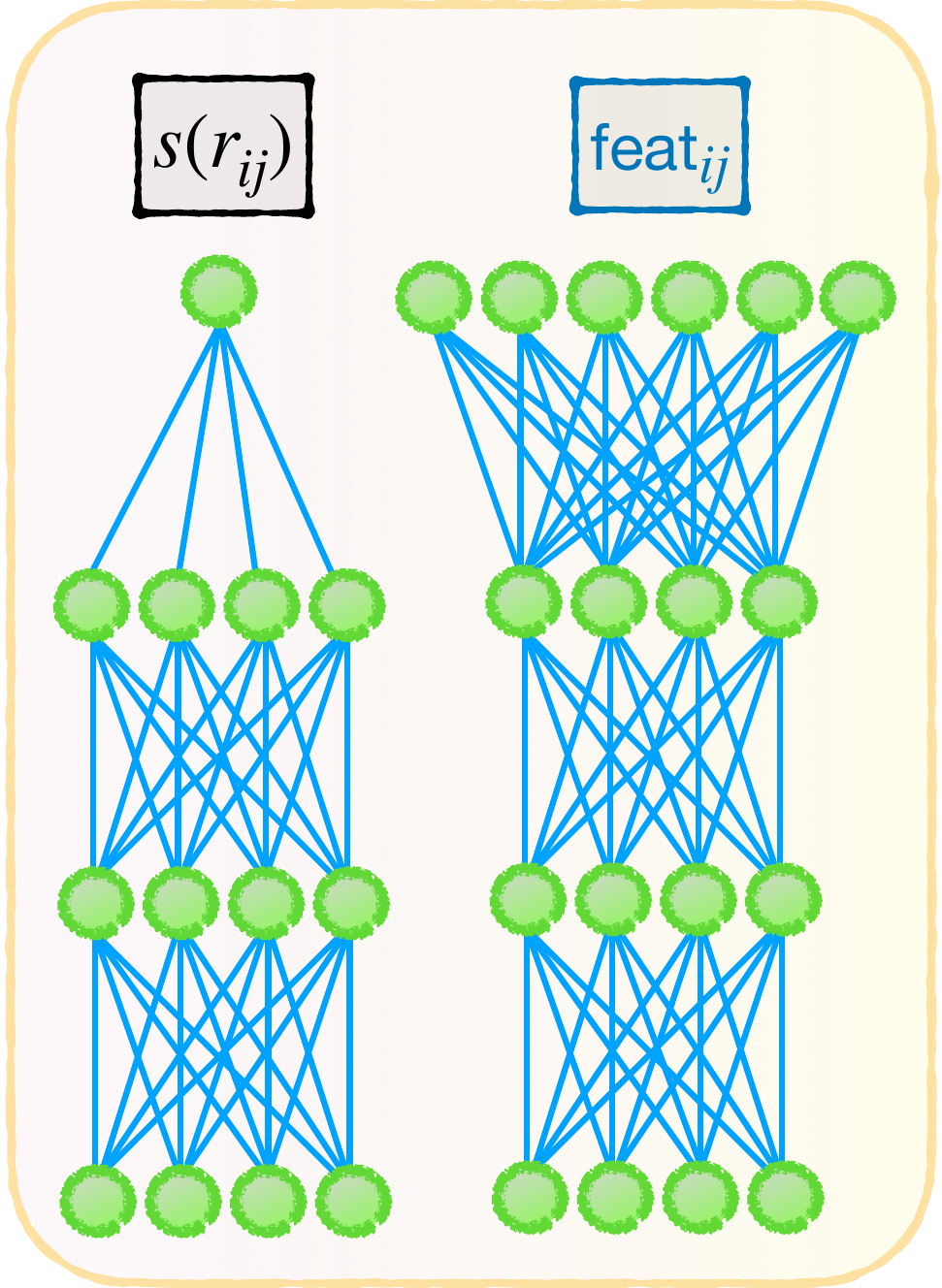}
     \end{subfigure}
     \caption{Illustration of the message-passing mechanism. Left: The features are aggregated from neighboring features repeatedly, effectively increasing the cutoff radius of the model. Right: Comparing the embedding network of the first pass and the MP pass, with the input of the latter being a multidimensional feature associated to an atom pair $i,j$. }
     \label{fig:mp_plot}
\end{figure}

Now we describe our enhanced DP model with message passing, or DP-MP. The architecture is illustrated in Figure~\ref{fig:mp} and \ref{fig:mp_plot}. Message passing is a common design in Graph Neural Networks, which allows one to learn feature on top of features iteratively. When applied to the DP model, the idea is simple: After computing the embedding network and summing over the neighbors, we obtain per-atom features $T_i^{(1)}, T_i^{(3)}$ and $D_i$ from step 3 and 4. This can be used as a starting point for a new round of embedding calculations for each neighbor pair $(i,j)$. The only difference is that, for the first round of embedding network in step 2, the input is only a scalar $s(r_{ij})$. But now we have much more information related to a pair $(i,j)$ like
$\bfr_{ij}, G(s(r_{ij})), T_i, D_i, T_j, D_j$ and so on. 
Among them, we make use of the invariant features $G(s(r_{ij})), D_i, D_j$, as well as create a new set of invariant features: $\langle T_i^{(3)}, \bfr_{ij}\rangle_3$ and $\langle T_j^{(3)}, \bfr_{ij}\rangle_3$. These invariant features are concatenated 
\begin{align*}
    \text{feat}_{ij}=\text{concatenate} \bracket{G(s(r_{ij})), D_i, D_j, \langle T_i^{(3)}, \bfr_{ij}\rangle_3, \langle T_j^{(3)}, \bfr_{ij}\rangle_3}
\end{align*}
as the new input to the embedding network.

This process can be iterated: After each embedding network pass, we aggregate features from neighbors by step 3. We obtain new $T$ and $D$ atomic features from step 3 and 4, which are used in the input to the new embedding pass starting from step 2. After a few loops we can terminate and enter the previous fitting process described in step 5.

At the final loop at step 3, a slightly different feature set is employed: Instead of $T_i^{(1)}$ and $T_i^{(3)}$, we use $T_i^{(3)}$ and $T_i^{(6)}$, where $T_i^{(6)}$ is a set of $6$-vectors defined by
\[ 
T_i^{(6)} = \frac{1}{N_{nbr}}\sum_{j \in \mathcal{N}_i} s(r_{ij})[\hat{\bfr}_{ij}]_6 G(s(r_{ij})).
\]
Here $[x]_6 = (x^2, y^2, z^2, \sqrt{2}xy, \sqrt{2}xy, \sqrt{2}xy)$, a $6$-vector that incorporates 0-th and 2-nd order tensorial information. The subsequent step 4 is computed by
\[ 
    D_{i,(m_1, m_2)} = \langle T_{i,m_1}^{(3)}, T_{i,m_2}^{(3)}\rangle_3 + \langle T_{i,m_1}^{(6)}, T_{i,m_2}^{(6)}\rangle_6,
\]
still invariant under rotation. We find this to be a good balance between improving the expressive power of the model and not incurring much computational cost. In fact, these $T$ features can be mathematically interpreted as a subset of the complete equivariant representations\cite{batzner20223}.

There are certain implementation details that we did not dive into for the sake of clarity. Firstly, in calculating $s(r_{ij})$, we use a slightly simpler function than the original DP
\[ 
s(r) = \begin{cases}
    \frac{1}{r} \bracket{1-3\bracket{\frac{r}{r_c}}^2 + 2\bracket{\frac{r}{r_c}}^3} &\text{ if } r < r_c, \\
    0 &\text{ if } r \ge r_c.
\end{cases}
\]
And the calculated $s(r)$ is shifted and normalized on a per-atom-type basis to have zero mean and unit variance before entering the embedding network, with the same normalizing factor (but no shift) applied in the $s(r)$ as well in the summations of step 3. Secondly, the linear transformation of the first layer in the MP embedding net is actually performed on $T_i^{(3)}$ and $T_j^{(3)}$ before the inner product with $\hat{\bfr}_{ij}$ and concatenation in feat$_{ij}$, which gives the equivalent math with reduced computational cost. For more details, we refer to the published code.

\subsection{Iterative Wannier Center Prediction}
\label{sec:DW}

Maximally localized Wannier functions give a well-defined alternative representation of the Bloch wave functions for the valence electrons in insulators. 
They are localized in space, and their distributional centers, short as \emph{Wannier centers} (WCs), can be seen as representing the centers of charge of individual electrons.
WCs are connected to the local and global polarization of the system\cite{marzari2012maximally, zhang2020deep}.
They are also used to explicitly calculate the long-range dipole-dipole Coulomb interactions\cite{zhang2022deep}, which is important in the study of charged systems.

The previous DP model has been used to predict the \emph{Wannier centroid}, defined as the average position of WCs associated with a certain atom.
For example, in an \ce{H2O} molecule, there are 4 WCs associated with it.
Each one of them represents a pair of electrons with opposite spin, with two of them for the bonding pairs and two for the lone pairs.
The dipole moment is determined by the average of the 4 WCs or the Wannier centroid.
The Wannier centroid obtained from DFT calculations can be learned by a separate neural network in DP, sometimes called the Deep Wannier (DW) model\cite{zhang2020deep}.
Compared to the standard DP model which predicts a scalar energy for each atom and then sums them up, DW predicts a vector for each Oxygen atom, representing its relative position to the Wannier centroid. 
This is achieved by modifying the final step 5 where one changes the fitting network's output $F(D_i)$ to be a feature of length $M_1$, and the relative displacement from the $i$-th Oxygen atom is expressed by $\langle F(D_i), T_i^{(3)}\rangle_{M_1}$.

DW works well on predicting Wannier centroids in water, or more generally, insulating systems
where you can assign the WCs to individual atoms. However, in general, WCs are not unambiguously associated with certain atoms.
Examples include \ce{e-}(aq), as well as more complex reactions involving electron transfer.
Still, WCs are functions of the atomic coordinates under the Born-Oppenheimer approximation.
This calls for a new scheme to predict the WCs without anchoring them to given atoms.

\begin{figure}[h]
          \centering
          \includegraphics[clip, trim=8cm 14.1cm 16.5cm 11.7cm, width=0.45\textwidth]{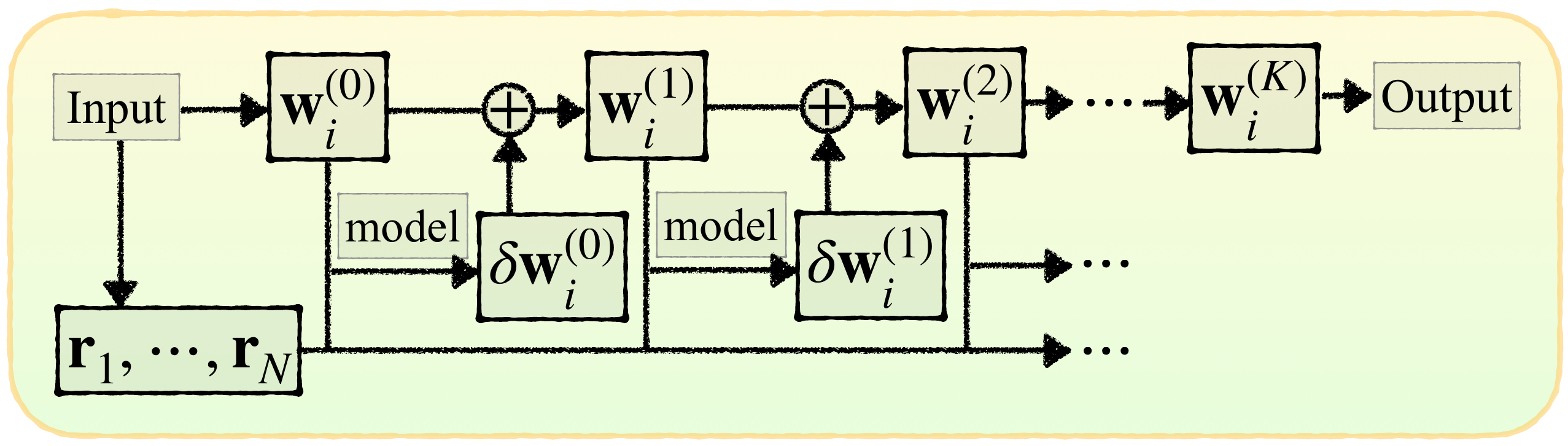}
         \caption{Illustration of DWIR, where the model takes the current WC prediction as part of the input and predicts an update.}
         \label{fig:DWIR}
\end{figure}

Here we introduce the new Deep Wannier Iterative Refinement (DWIR) model. 
The idea is simple: Now the WCs are anchored to themselves, and we predict only a correction displacement on top of a given prediction. 
Suppose the atomic coordinates are $\{\bfr_i\}_{i=1}^N$, and we have some initial guess of the WCs $\{\bfw_j^{(0)}\}_{j=1}^{N_w}$, where $N_w$ is the total number of WCs.
The initial guess is subject to errors, but we use a DW-like model to correct it iteratively:
\[ \delta \bfw_j^{(k)} = \text{model}\bracket{\bfr_1, \cdots, \bfr_N, \bfw_1^{(k)},\cdots, \bfw_{N_w}^{(k)}}_j \]
  and 
\[ \bfw_j^{(k+1)} = \bfw_j^{(k)} + \delta \bfw_j^{(k)}. \]
Starting from $k=0$, the model is reused to iterate $K$ times, and we obtain the final prediction $\{\bfw_j^{(K)}\}_{j=1}^{N_w}$. We train the model with a loss function
\[ L = \sum_{k=1}^{K}\gamma^k \ell(\bfw^{(k)}, \bfw^\ast)\]
where $\ell$ is a loss function for an individual prediction, $\gamma > 1$ is some fixed constant, and $\bfw^\ast$ stands for the true WCs, whose permutation is determined by a greedy pairing with the predicted WCs based on a closest-distance principle. The scheme penalizes errors at later iterations, encouraging the process to converge to a fixed point that equals the true WCs in just a few iterations. 

In a model's architecture, the WCs are treated just as a different kind of point particle, so any existing model can be used here, such as using DP-MP for improved accuracy. 
Also, DWIR can work with either spin-saturated or spin-polarized calculations.
The latter is used in the \ce{e-}(aq) system where one WC represents one electron instead of a pair.

The initial guess, while not important for the final result, should not deviate too much from the true WCs, otherwise the model will have a hard time converging.
For example, in water, one can initialize 4/8 (spin-saturated/spin-polarized) random WCs around each Oxygen atom during training.
In \ce{e-}(aq), one can initialize the excess electron's WC to be within 1~\AA\ of the true WC during training.
In a simulation, one simply uses the previous step's prediction as the initial guess for the next step.

Again, there are certain implementation details.
For example, $s(r)$ is modified to be finite at $r=0$ to handle potentially overlapping particle positions.
For more details, we refer to the published code.

\section{Benchmark Results on Water}

In this section, we present a simple benchmark result of our enhanced models on a water system.
Our dataset consists of some short DFT simulation trajectories of a periodic box of 128 \ce{H2O} molecules totaling a few picoseconds,
which are split into a training set of 7797 configurations and a validation set of 1501 configurations.
We use the same DFT functional as in the \ce{e-}(aq) simulation (described in Sec.~\ref{sec:sim}) apart from doing a spin-saturated calculation without the excess electron.

We use a cutoff radius of $6$~\AA\, and find that one MP pass in the DP-MP model is best in achieving a good accuracy while offering a significant speed advantage over other models.
We use the default network width (number of neurons in each layer) of (32,32) for the initial embedding network, (64,32,64) for the MP embedding network, and (64,64,64,1) for the fitting network.
We also benchmark the newly implemented DP model (referred to as DP(JAX)) with the default embedding network width (32,32,64).\footnote{
The embedding network of the DP model, as well as the initial embedding network of the DP-MP model, is compressed by default.
}
Apart from the implementation, it differs from the original DP\cite{zeng2023deepmd}(referred to as DP(TF)) in that
the per-atom features in Step 3 is ($T_i^{(3)}$, $T_i^{(6)}$) as in the final loop of in DP-MP.

\begin{figure}[h]
          \centering
          \includegraphics[clip, trim=0cm 0cm 0cm 0cm, width=0.33\textwidth]{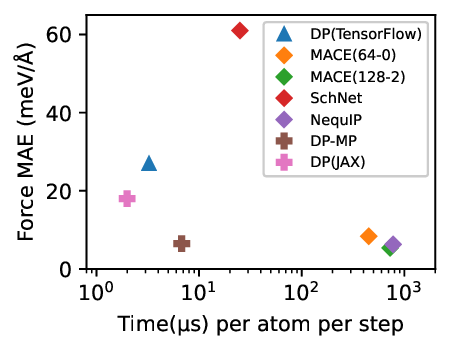}
         \caption{Accuracy-speed trade-off on the PBEh(0.4) water system.}
         \label{fig:tradeoff}
\end{figure}

We measure the root mean square error (RMSE) or mean absolute error (MAE) of force predictions on the validation set.
We also measure the speed of the models in simulations of a system of 128 \ce{H2O} molecules on a single NVIDIA A100 GPU.
The results are summarized in Table~\ref{tab:water}. 
\begin{table}[h]
    \centering
    \caption{Comparison of DP models on the PBEh(0.4) water system. Units are meV/\AA\ for force RMSE and MAE, and $\mu$s/atom/step for computational cost.}
    \label{tab:water}
    \begin{tabular}{l|c|c|c}
    \hline
    Model & Force RMSE & Force MAE & Cost \\
    \hline
    DP(TF) & 35 & 27 & 3.25 \\
    DP(JAX) & 23 & 18 & 1.99 \\
    DP-MP & 9.3 & 6.5 & 6.77 \\
    \hline
    \end{tabular}
\end{table}
We also compare with other neural network force fields including
SchNet\cite{schutt2017schnet}, NequIP\cite{liao2022equiformer}, and MACE\cite{Batatia2022mace}.
\footnote{We employ the training parameters of SchNet and NequIP from \citet{fu2022forces} and MACE from the official documentation with a small(64-0) and large(128-2) model.
Since our dataset is larger than the examples in these references
we reduce the number of training epochs accordingly but ensure that further training does not improve the validation error.}
\footnote{While there are presumably more accurate models, they tend to be even slower and are not included in the comparison.}
We plot the accuracy-speed trade-off in Figure~\ref{fig:tradeoff}.
It can be seen that DP-MP is almost two orders of magnitude faster compared to other models with similar accuracy.
This is largely due to the design of the model itself, but various implementation gains
\footnote{Firstly, JAX tends to be somewhat faster than other packages like PyTorch or TensorFlow with which the other models are implemented.
Also, we connect the model with JAX-MD, enabling an end-to-end GPU workflow.
The simulation of other models either uses the LAMMPS(DP(TF)) or ASE(SchNet, NequIP, MACE) interface.
While the simulation part is not the computational bottleneck compared to the evaluation of the neural network,
such an interface can still cause some overhead.
In addition, we use 32-bit floating point accuracy in DP-MP and DP(JAX) by default, which we find to have no impact on the prediction accuracy.} play an important role as well.
Together, the new enhanced models (DP-MP and DP(JAX)) achieve a great balance between accuracy and speed and offer a good choice for large-scale simulations.

\section{Simulating the Solvated Electron in Water}
\label{sec:sim}

The solvated electron in water, also called the hydrated electron, is a byproduct of water radiolysis, a simple and potent reducing agent, and the culprit for DNA damage in biological systems. It has been attracting interest for decades of studies\cite{hart1962absorption, schnitker1987quantum, turi2005characterization, turi2012theoretical, boero2003first, larsen2010does, uhlig2012unraveling, savolainen2014direct, uhlig2014optical, alizadeh2015biomolecular, herbert2017hydrated, ambrosio2017electronic, herbert2019structure, svoboda2020real, lan2021simulating}.
Upon being created by ionizing radiation, it occupies a delocalized state as a quasi-free electron.
Then, on a picosecond timescale, it thermalizes by creating a cavity in the surrounding water molecules, and localizes into a stable state. It is now generally agreed that the localized state is a quasi-spherical cavity model with a shell of surrounding water molecules\cite{herbert2017hydrated, ambrosio2017electronic}.

We will only focus on studying the localized state.
One reason is that non-adiabatic effects can be present in the delocalized state, which are not captured by electronic ground state simulations.
Another reason would be that ML models are agnostic to the total number of electrons.
If a model were to be transferable to larger systems, it should work for both \ce{e-}(aq) and normal bulk water.
Upon creating a delocalized electron in a finite box, the atoms are still at a bulk water configuration, indistinguishable to the ML model,
but the atomic forces become different, making the forces ill-defined if these states are to be included.

\begin{figure}[h]
          \centering
          \includegraphics[clip, trim=15cm 4cm 21cm 3cm, width=0.49\textwidth]{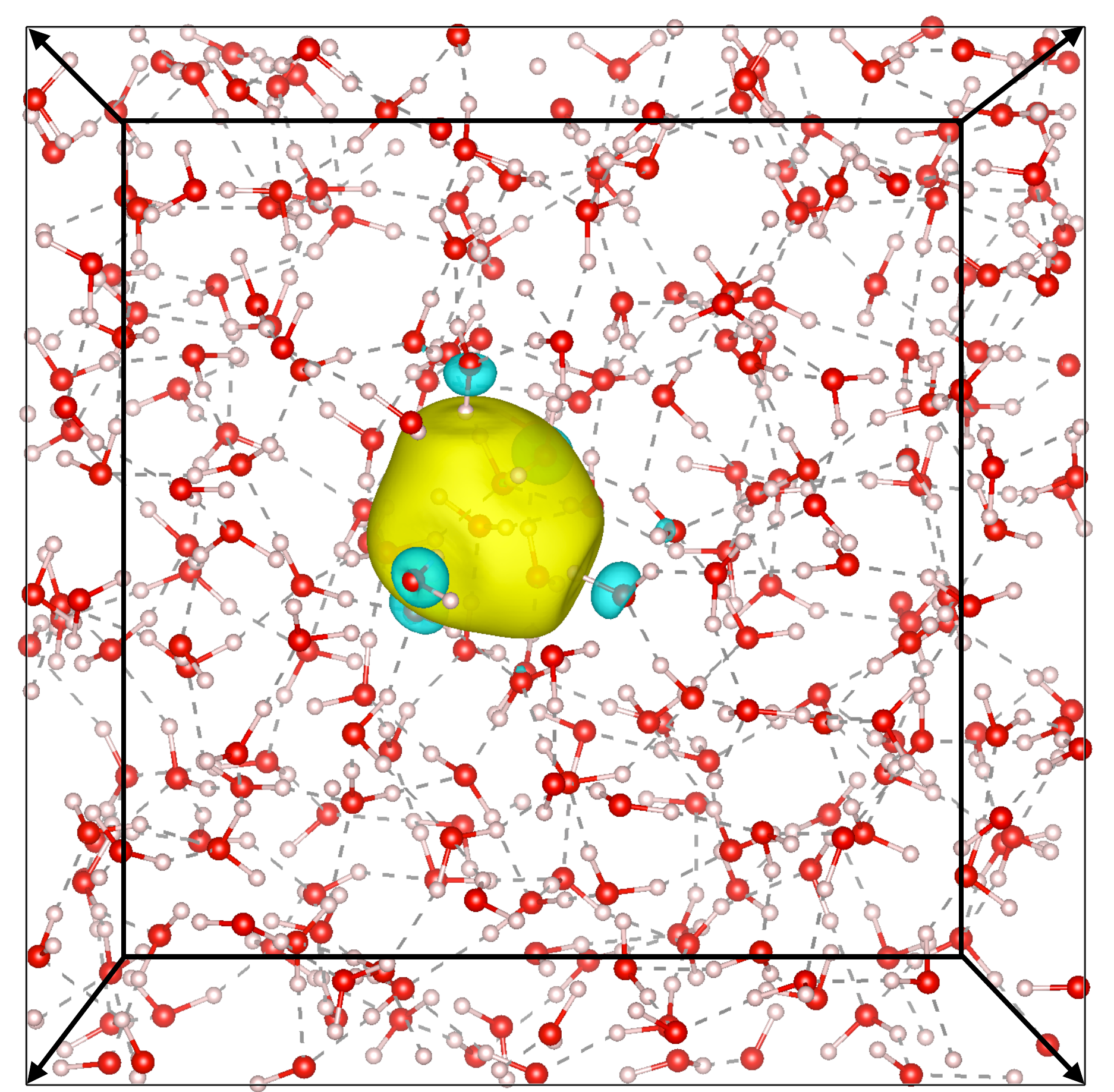}
         \caption{Picture of the positive and negative isosurfaces of the Wannierized wave function of \ce{e-}(aq). Surrounding water molecules point toward the electron through one hydrogen atom, resembling an H-bond. The solvated electron causes disturbances to the H-bond network in water. The two boxes represent training on a smaller system and transferring to a larger system. }
         \label{fig:polaron}
\end{figure}

\begin{figure*}[!ht]
\centering
     \begin{subfigure}[b]{0.4\textwidth}
    \caption{}
         \includegraphics[height=5cm]{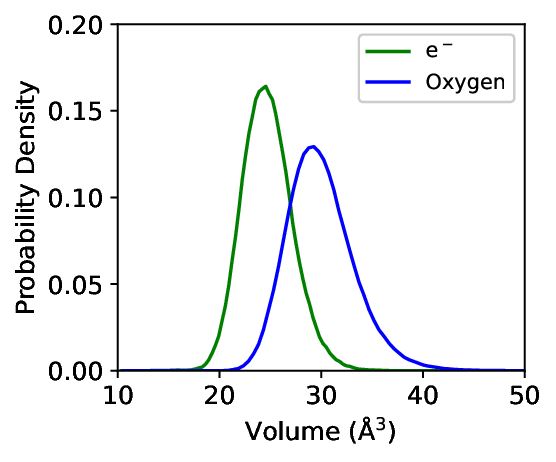}
         \label{fig:voronoi}
     \end{subfigure}
     \begin{subfigure}[b]{0.4\textwidth}
    \caption{}
          \includegraphics[height=5cm]{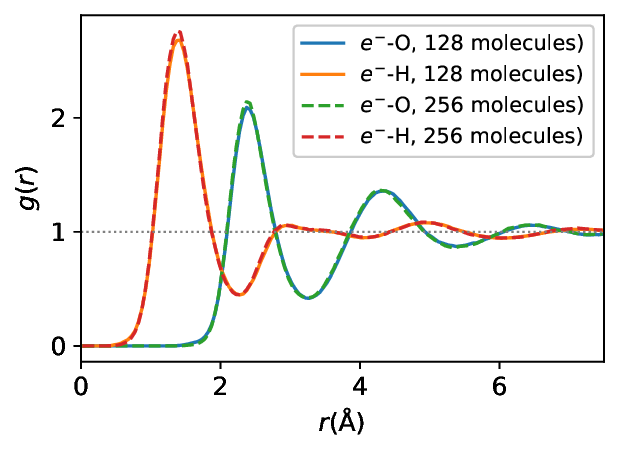}
         \label{fig:rdf}
     \end{subfigure}
     \\
    \begin{subfigure}[b]{0.32\textwidth}
    \caption{}
          \includegraphics[height=4.5cm]{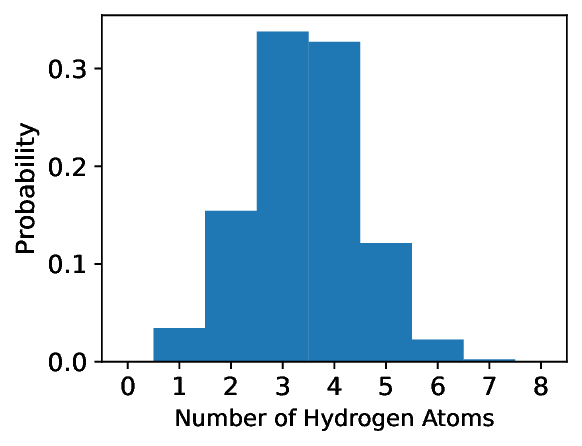}
         \label{fig:nh}
     \end{subfigure}
     \begin{subfigure}[b]{0.32\textwidth}
     \caption{}
          \includegraphics[height=4.5cm, clip, trim=0cm 0cm 0cm 0.04cm]{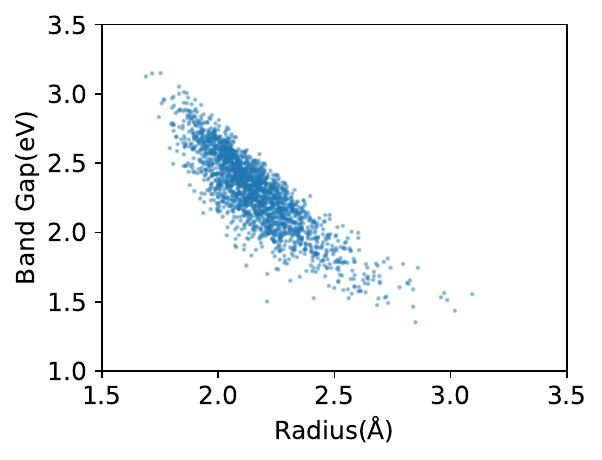}
         \label{fig:radius}
     \end{subfigure}
     \begin{subfigure}[b]{0.32\textwidth}
     \caption{}
          \includegraphics[height=4.5cm]{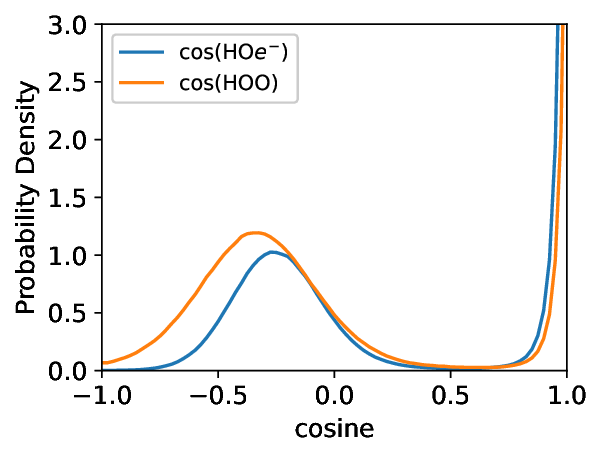}
         \label{fig:cos}
     \end{subfigure}
    \caption{ Results on the structure of \ce{e-}(aq). (a): Distribution of the Voronoi volume of \ce{e-} and water molecules. (b): Radial distribution function between \ce{e-} and O/H atoms.  (c) Distribution of the number of hydrogen atoms in the first shell of \ce{e-}. (d) Radius of gyration of the Wannierized state of \ce{e-}, plotted against the Kohn-Sham band gap. (e) Distribution of the cosine angle of H-O-\ce{e-} and H-O-O, showing the resemblance of the H-\ce{e-} bond to the H-bond.}
    \label{fig:result}
    \end{figure*}

\subsection{Setup for DFT Simulation}

We first perform a DFT simulation in a periodic box of 128 \ce{H2O} molecules plus one \ce{e-}, with the NVT ensemble at experimental density.
We use the CP2K software\cite{kuhne2020cp2k} and adopt a setting described as follows, which has been used in previous works and well-tested for the description of \ce{e-}(aq)\cite{ambrosio2016structural, ambrosio2017electronic, pizzochero2019picture}.
We use the hybrid functional PBEh($\alpha$), with the fraction of Fock exchange $\alpha$ set to 0.4.
The van der Waals correction is included by the rVV10 functional where the parameter $b$ is set to 5.3.
We use the triple-$\zeta$ polarized (TZP) basis and Goedecker–Teter–Hutter pseudopotentials.
The charge density, expanded in a plane-wave basis, has an 800~Ry cutoff.
We use a spin-polarized calculation with an added uniform background charge to neutralize the system.
The temperature is maintained at 350 K via the use of a Nosé-Hoover thermostat in order to ensure a frank diffusive motion\cite{ambrosio2017electronic}.
Starting from the initial equilibrated bulk water configuration, it takes around 0.2$\sim$0.4 ps for the excess electron to localize, and the initial configurations that are not fully localized are not used in model training.

\subsection{Setup for Model Training and Simulation}

We train a DP-MP model to simulate the solvated electron.
We use a cutoff radius of $6$~\AA, same as previous DP models for bulk water, and employ a single MP pass.
The loss function is a sum of energy and force terms similar to that used in the training of DP.
The model is trained with a batch size of 1 for 500,000 batches using the Adam optimizer with an exponentially decaying learning rate from $2\times10^{-3}$ to $10^{-6}$.
The training takes around 1.5 hours on an NVIDIA A100 GPU.
An active learning procedure, DP-GEN\cite{zhang2020dp}, is followed to improve the model's robustness.
This involves training several initial models with different random seeds, performing a simulation with one of the trained models,
and sampling a small set of extra configurations from the trajectory based on a model deviation metric.
These extra configurations are then labeled by DFT calculations and added to the training set.
The initial AIMD trajectories have a total length of around $15$ ps.
With some extra configurations from DP-GEN, the final training set has around 30,000 configurations.

To keep track of the electron's position, we also train a DWIR model.
We use the same DP-MP base architecture. 
The configurations used for training are also the same, with the WCs calculated from the Kohn-Sham orbitals of the DFT calculations.
Since we're only interested in the excess electron here, only one WC per configuration needs to be predicted by the model, 
though the DWIR model can equally well keep track of all the electrons.
We perform $K=4$ iterations of refinement.
We use a batch size of 64 and trained for 50,000 batches using Adam with an exponentially decaying learning rate from $10^{-2}$ to $10^{-4}$.

The DP-MP model achieves a root mean square validation error of 12~meV/\AA\ for the forces.
The DWIR model achieves a root mean square validation error of 0.025~\AA\ for the WC.
The DP-MP model is then used to perform a 1 ns-long simulation\footnote{In practice, the lifetime of \ce{e-}(aq) may not be this long
 due to the reaction with other species like the hydronium ion,
 but our simulation aims at collecting the statistics of  \ce{e-}(aq) itself.
}
in the same NVT ensemble as the DFT simulation.
To show the transferability of the model, we also perform a simulation of a larger system of 256 \ce{H2O} molecules plus one \ce{e-}, with the same DP-MP model.
The DWIR model is used to predict the WCs after the simulation, and the WCs are then used to calculate various properties of the solvated electron.

\subsection{Results on the Solvated Electron}

Our simulation confirms the cavity model,
where the electron has been observed to remain stably localized as depicted in Figure~\ref{fig:polaron},
both in DFT and DP-MP simulations.
The structural results are shown in Figure~\ref{fig:result}. Our results are in general consistent with the DFT results from the previous literature\cite{ambrosio2017electronic}, but certain sensitive numbers vary because DP-MP based long trajectories give more converged statistics than previous AIMD trajectories.

We first conduct a Voronoi analysis on the WC of \ce{e-} and all oxygen atoms.
The respective volume distributions are shown in Figure~\ref{fig:voronoi}.
The volume of \ce{e-} is smaller than that occupied by a water molecule.
From the average volume we deduce a radius of 1.80~\AA\, compared to 1.92~\AA\ for a water molecule.

In Figure~\ref{fig:rdf}, we show the radial distribution function between the WC of \ce{e-} and O/H atoms.
The first peak is at 1.4~\AA\ for \ce{e-}-H, and 2.4~\AA\ for \ce{e-}-O.
The first minimum is at 2.3~\AA\ for \ce{e-}-H, and 3.3~\AA\ for \ce{e-}-O. We obtain the result for the 256-molecule system as well. It is slightly more structured and localized, which alludes to the importance of using a large enough box to simulate $\ce{e-}(aq)$, where a smaller box lowers the energy barrier for delocalization. But the difference between system size 128 and 256 is already tiny, indicating valid results with 128 molecules.

We compute the coordination number of hydrogen atoms within the first minimum 2.3~\AA,
resulting in a mean of 3.4\footnote{This result is smaller than previous results \cite{ambrosio2017electronic} because the coordination number is quite sensitive to the measurement of the minimum of the radial distribution function, where we give a slightly smaller 2.3~\AA.}  and a high standard deviation of 1.1, where the distribution spans from 1 to 7 as shown in Figure~\ref{fig:nh}.
This is due to a highly volatile H-bond network in the vicinity of the electron.

By sampling 2000 configurations from the trajectory and conducting DFT calculations again, we compute the radius of gyration from the spread of the Wannierized wave function of \ce{e-}.
The average number is 2.16~\AA\footnote{This radius is slightly smaller than previous estimates\cite{ambrosio2017electronic} based on the Bloch state, but gives the same qualitative picture since the state of \ce{e-} does not mix much with the valence band during Wannierization.} with a standard deviation of 0.18~\AA, which, compared to the radius inferred from the Voronoi volume,
indicates that the electronic density extends into the first shell. 
It is still localized but much less localized than the Wannierized valence-band electrons in water. 
The radius of gyration is plotted against the Kohn-Sham band gap in Figure~\ref{fig:radius}, 
showing a clear negative correlation, where a larger radius corresponds to a smaller band gap and a state of higher energy. A lack of an extended tail at the bottom-right is an indication of stable localization. 

To examine the nature of the interaction between \ce{e-} and surrounding water molecules,
we compute the distribution of the cosine angle of H-O-\ce{e-}, conditioning on H-O being covalently bonded, as well as the distance between
O and \ce{e-} being within 3.0~\AA.
This is compared with the angle of H-O-O, or H-O$_1$-O$_2$, where H-O$_1$ is covalently bonded and O$_1$-O$_2$ is within 3.0~\AA. 
The latter shows a strong peak at 0\textdegree, which corresponds to the H-bond angle, with another soft peak that that stands for the other H atom covalently bonded to O$_1$.
The cosine of H-O-\ce{e-} also shows the similar two peaks, which indicates that H-\ce{e-} bonds are similar to H-bonds,
with one hydroxyl group pointing to the electron as indicated in Figure~\ref{fig:polaron}.
The soft peak for water is at around \mbox{-0.33}, corresponding to a tetrahedral H-bond network, while the soft peak for \ce{e-} is at around \mbox{-0.27}, much closer to the cosine angle of the water molecule itself. This indicates that the H-bond network is disturbed by the excess electron.

To understand the diffusion properties of \ce{e-}(aq), additional NVE simulations are performed at the same temperature.
By fitting the mean square displacement to the Einstein relation, the diffusion coefficient of \ce{e^-}(aq) is calculated as $0.33\pm0.01$~\AA$^2$/ps.
As a comparison, the diffusion coefficient for bulk water molecules under the same setting is $0.24\pm0.01$~\AA$^2$/ps.
The absolute value of the diffusion coefficient depends on various factors like the DFT functional and the system size,
but the relative value indicates that the solvated electron is more mobile than water molecules.
The fast diffusion is the result of the frequent entry and exit of water molecules into the shell surrounding the electron. The solvated electron acts as a H-bond acceptor but not a donor, 
disturbing the H-bond network in water.
This is already implied by the high variance of the \ce{e-}-H coordination number.
To see it more, we calculate the survival time of the hydogen bonds as well as the H-\ce{e-} bond, defined by a unified geometric criterion due to their resemblance:
The distance between the donor(O) and acceptor(O or \ce{e-}) is less than 3.3~\AA, and the H-O-O/H-O-\ce{e-} angle is less than 30\textdegree.
The bond is deemed broken if the H atom forms a different bond according to this criterion, or a weaker criterion of 3.6~\AA/60\textdegree\ is violated.
The average survival time is calculated to be 0.81 ps for H-bonds, and 0.58~ps for H-\ce{e-} bonds.
The relative value directly indicates that the H-\ce{e-} bond is less stable than H-bonds, consistent with the increased diffusion.

\section{Conclusion}

In this work, we have enhanced the design of Deep Potential models, providing an excellent balance between accuracy and speed.
We have also introduced a new DWIR model to predict Wannier centers without relying on atom anchoring.
These models have been applied to simulate the solvated electron in water, providing new insights into the structure and dynamics of the system.
We expect the new DP and DP-MP models to set a new standard for the simulation of complex systems.
We also expect our models to be a useful tool to simulate more complex electron transfer reactions in future work.

\section*{Acknowledgements}

We acknowledge Yixiao Chen, Linfeng Zhang, and Duo Zhang for very useful discussions during the development of the model. This work is supported by the Computational Chemical Sciences Center “Chemistry in Solution and at Interfaces” funded by the U.S. Department of Energy under Award No.~DE-SC0019394. This research mainly uses resources of the National Energy Research Scientific Computing Center (NERSC), which is supported by the U.S. Department of Energy (DOE), Office of Science under Contract No.~DE-AC02-05CH11231. We also acknowledge Princeton Research Computing at Princeton University for providing resources.

\renewcommand\refname{References}

\bibliography{reference} 
\bibliographystyle{ieeetr} 
\end{document}

%% file: math_commands.tex
\newcommand{\bracket}[1]{\left(#1\right)}

\newcommand{\bfw}{\mathbf{w}}

\newcommand{\bfr}{\mathbf{r}}

%% file: main.bbl
\begin{thebibliography}{10}

\bibitem{behler2007generalized}
J.~Behler and M.~Parrinello, ``Generalized neural-network representation of high-dimensional potential-energy surfaces,'' {\em Physical review letters}, vol.~98, no.~14, p.~146401, 2007.

\bibitem{schutt2017quantum}
K.~T. Sch{\"u}tt, F.~Arbabzadah, S.~Chmiela, K.~R. M{\"u}ller, and A.~Tkatchenko, ``Quantum-chemical insights from deep tensor neural networks,'' {\em Nature communications}, vol.~8, no.~1, p.~13890, 2017.

\bibitem{gilmer2017neural}
J.~Gilmer, S.~S. Schoenholz, P.~F. Riley, O.~Vinyals, and G.~E. Dahl, ``Neural message passing for quantum chemistry,'' in {\em International conference on machine learning}, pp.~1263--1272, PMLR, 2017.

\bibitem{han2017deep}
J.~Han, L.~Zhang, R.~Car, {\em et~al.}, ``Deep potential: A general representation of a many-body potential energy surface,'' {\em arXiv preprint arXiv:1707.01478}, 2017.

\bibitem{gasteiger2020directional}
J.~Gasteiger, J.~Gro{\ss}, and S.~G{\"u}nnemann, ``Directional message passing for molecular graphs,'' {\em arXiv preprint arXiv:2003.03123}, 2020.

\bibitem{hu2021forcenet}
W.~Hu, M.~Shuaibi, A.~Das, S.~Goyal, A.~Sriram, J.~Leskovec, D.~Parikh, and C.~L. Zitnick, ``Forcenet: A graph neural network for large-scale quantum calculations,'' {\em arXiv preprint arXiv:2103.01436}, 2021.

\bibitem{gasteiger2021gemnet}
J.~Gasteiger, F.~Becker, and S.~G{\"u}nnemann, ``Gemnet: Universal directional graph neural networks for molecules,'' {\em Advances in Neural Information Processing Systems}, vol.~34, pp.~6790--6802, 2021.

\bibitem{ying2021transformers}
C.~Ying, T.~Cai, S.~Luo, S.~Zheng, G.~Ke, D.~He, Y.~Shen, and T.-Y. Liu, ``Do transformers really perform badly for graph representation?,'' {\em Advances in Neural Information Processing Systems}, vol.~34, pp.~28877--28888, 2021.

\bibitem{liu2022spherical}
Y.~Liu, L.~Wang, M.~Liu, Y.~Lin, X.~Zhang, B.~Oztekin, and S.~Ji, ``Spherical message passing for 3d molecular graphs,'' in {\em International Conference on Learning Representations (ICLR)}, 2022.

\bibitem{batzner20223}
S.~Batzner, A.~Musaelian, L.~Sun, M.~Geiger, J.~P. Mailoa, M.~Kornbluth, N.~Molinari, T.~E. Smidt, and B.~Kozinsky, ``E (3)-equivariant graph neural networks for data-efficient and accurate interatomic potentials,'' {\em Nature communications}, vol.~13, no.~1, p.~2453, 2022.

\bibitem{liao2022equiformer}
Y.-L. Liao and T.~Smidt, ``Equiformer: Equivariant graph attention transformer for 3d atomistic graphs,'' in {\em The Eleventh International Conference on Learning Representations}, 2022.

\bibitem{zhang2018deep}
L.~Zhang, J.~Han, H.~Wang, R.~Car, and E.~Weinan, ``Deep potential molecular dynamics: a scalable model with the accuracy of quantum mechanics,'' {\em Physical review letters}, vol.~120, no.~14, p.~143001, 2018.

\bibitem{zhang2018end}
L.~Zhang, J.~Han, H.~Wang, W.~Saidi, R.~Car, {\em et~al.}, ``End-to-end symmetry preserving inter-atomic potential energy model for finite and extended systems,'' {\em Advances in neural information processing systems}, vol.~31, 2018.

\bibitem{jia2020pushing}
W.~Jia, H.~Wang, M.~Chen, D.~Lu, L.~Lin, R.~Car, E.~Weinan, and L.~Zhang, ``Pushing the limit of molecular dynamics with ab initio accuracy to 100 million atoms with machine learning,'' in {\em SC20: International conference for high performance computing, networking, storage and analysis}, pp.~1--14, IEEE, 2020.

\bibitem{lu202186}
D.~Lu, H.~Wang, M.~Chen, L.~Lin, R.~Car, E.~Weinan, W.~Jia, and L.~Zhang, ``86 pflops deep potential molecular dynamics simulation of 100 million atoms with ab initio accuracy,'' {\em Computer Physics Communications}, vol.~259, p.~107624, 2021.

\bibitem{jax2018github}
J.~Bradbury, R.~Frostig, P.~Hawkins, M.~J. Johnson, C.~Leary, D.~Maclaurin, G.~Necula, A.~Paszke, J.~Vander{P}las, S.~Wanderman-{M}ilne, and Q.~Zhang, ``{JAX}: composable transformations of {P}ython+{N}um{P}y programs,'' 2018.

\bibitem{schoenholz2019jax}
S.~S. Schoenholz and E.~D. Cubuk, ``Jax md: End-to-end differentiable, hardware accelerated, molecular dynamics in pure python,'' 2019.

\bibitem{marzari2012maximally}
N.~Marzari, A.~A. Mostofi, J.~R. Yates, I.~Souza, and D.~Vanderbilt, ``Maximally localized wannier functions: Theory and applications,'' {\em Reviews of Modern Physics}, vol.~84, no.~4, p.~1419, 2012.

\bibitem{zhang2020deep}
L.~Zhang, M.~Chen, X.~Wu, H.~Wang, E.~Weinan, and R.~Car, ``Deep neural network for the dielectric response of insulators,'' {\em Physical Review B}, vol.~102, no.~4, p.~041121, 2020.

\bibitem{herbert2017hydrated}
J.~M. Herbert and M.~P. Coons, ``The hydrated electron,'' {\em Annual review of physical chemistry}, vol.~68, pp.~447--472, 2017.

\bibitem{ambrosio2017electronic}
F.~Ambrosio, G.~Miceli, and A.~Pasquarello, ``Electronic levels of excess electrons in liquid water,'' {\em The journal of physical chemistry letters}, vol.~8, no.~9, pp.~2055--2059, 2017.

\bibitem{lan2021simulating}
J.~Lan, V.~Kapil, P.~Gasparotto, M.~Ceriotti, M.~Iannuzzi, and V.~V. Rybkin, ``Simulating the ghost: quantum dynamics of the solvated electron,'' {\em Nature communications}, vol.~12, no.~1, p.~766, 2021.

\bibitem{lan2022temperature}
J.~Lan, V.~V. Rybkin, and A.~Pasquarello, ``Temperature dependent properties of the aqueous electron,'' {\em Angewandte Chemie International Edition}, vol.~61, no.~38, p.~e202209398, 2022.

\bibitem{ambrosio2016structural}
F.~Ambrosio, G.~Miceli, and A.~Pasquarello, ``Structural, dynamical, and electronic properties of liquid water: A hybrid functional study,'' {\em The Journal of Physical Chemistry B}, vol.~120, no.~30, pp.~7456--7470, 2016.

\bibitem{he2016deep}
K.~He, X.~Zhang, S.~Ren, and J.~Sun, ``Deep residual learning for image recognition,'' in {\em Proceedings of the IEEE conference on computer vision and pattern recognition}, pp.~770--778, 2016.

\bibitem{lu2022dp}
D.~Lu, W.~Jiang, Y.~Chen, L.~Zhang, W.~Jia, H.~Wang, and M.~Chen, ``Dp compress: A model compression scheme for generating efficient deep potential models,'' {\em Journal of chemical theory and computation}, vol.~18, no.~9, pp.~5559--5567, 2022.

\bibitem{bonati2018silicon}
L.~Bonati and M.~Parrinello, ``Silicon liquid structure and crystal nucleation from ab initio deep metadynamics,'' {\em Physical review letters}, vol.~121, no.~26, p.~265701, 2018.

\bibitem{gartner2020signatures}
T.~E. Gartner~III, L.~Zhang, P.~M. Piaggi, R.~Car, A.~Z. Panagiotopoulos, and P.~G. Debenedetti, ``Signatures of a liquid--liquid transition in an ab initio deep neural network model for water,'' {\em Proceedings of the National Academy of Sciences}, vol.~117, no.~42, pp.~26040--26046, 2020.

\bibitem{niu2020ab}
H.~Niu, L.~Bonati, P.~M. Piaggi, and M.~Parrinello, ``Ab initio phase diagram and nucleation of gallium,'' {\em Nature communications}, vol.~11, no.~1, p.~2654, 2020.

\bibitem{andrade2020free}
M.~F.~C. Andrade, H.-Y. Ko, L.~Zhang, R.~Car, and A.~Selloni, ``Free energy of proton transfer at the water--tio 2 interface from ab initio deep potential molecular dynamics,'' {\em Chemical Science}, vol.~11, no.~9, pp.~2335--2341, 2020.

\bibitem{zeng2020complex}
J.~Zeng, L.~Cao, M.~Xu, T.~Zhu, and J.~Z. Zhang, ``Complex reaction processes in combustion unraveled by neural network-based molecular dynamics simulation,'' {\em Nature communications}, vol.~11, no.~1, p.~5713, 2020.

\bibitem{zhang2021phase}
L.~Zhang, H.~Wang, R.~Car, and E.~Weinan, ``Phase diagram of a deep potential water model,'' {\em Physical review letters}, vol.~126, no.~23, p.~236001, 2021.

\bibitem{zhang2022deep}
L.~Zhang, H.~Wang, M.~C. Muniz, A.~Z. Panagiotopoulos, R.~Car, {\em et~al.}, ``A deep potential model with long-range electrostatic interactions,'' {\em The Journal of Chemical Physics}, vol.~156, no.~12, 2022.

\bibitem{zeng2023deepmd}
J.~Zeng, D.~Zhang, D.~Lu, P.~Mo, Z.~Li, Y.~Chen, M.~Rynik, L.~Huang, Z.~Li, S.~Shi, {\em et~al.}, ``Deepmd-kit v2: A software package for deep potential models,'' {\em arXiv preprint arXiv:2304.09409}, 2023.

\bibitem{schutt2017schnet}
K.~Sch{\"u}tt, P.-J. Kindermans, H.~E. Sauceda~Felix, S.~Chmiela, A.~Tkatchenko, and K.-R. M{\"u}ller, ``Schnet: A continuous-filter convolutional neural network for modeling quantum interactions,'' {\em Advances in neural information processing systems}, vol.~30, 2017.

\bibitem{Batatia2022mace}
I.~Batatia, D.~P. Kovacs, G.~N.~C. Simm, C.~Ortner, and G.~Csanyi, ``{MACE}: Higher order equivariant message passing neural networks for fast and accurate force fields,'' in {\em Advances in Neural Information Processing Systems} (A.~H. Oh, A.~Agarwal, D.~Belgrave, and K.~Cho, eds.), 2022.

\bibitem{fu2022forces}
X.~Fu, Z.~Wu, W.~Wang, T.~Xie, S.~Keten, R.~Gomez-Bombarelli, and T.~Jaakkola, ``Forces are not enough: Benchmark and critical evaluation for machine learning force fields with molecular simulations,'' {\em arXiv preprint arXiv:2210.07237}, 2022.

\bibitem{hart1962absorption}
E.~J. Hart and J.~W. Boag, ``Absorption spectrum of the hydrated electron in water and in aqueous solutions,'' {\em Journal of the American Chemical Society}, vol.~84, no.~21, pp.~4090--4095, 1962.

\bibitem{schnitker1987quantum}
J.~Schnitker and P.~J. Rossky, ``Quantum simulation study of the hydrated electron,'' {\em The Journal of chemical physics}, vol.~86, no.~6, pp.~3471--3485, 1987.

\bibitem{turi2005characterization}
L.~Turi, W.-S. Sheu, and P.~J. Rossky, ``Characterization of excess electrons in water-cluster anions by quantum simulations,'' {\em Science}, vol.~309, no.~5736, pp.~914--917, 2005.

\bibitem{turi2012theoretical}
L.~Turi and P.~J. Rossky, ``Theoretical studies of spectroscopy and dynamics of hydrated electrons,'' {\em Chemical reviews}, vol.~112, no.~11, pp.~5641--5674, 2012.

\bibitem{boero2003first}
M.~Boero, M.~Parrinello, K.~Terakura, T.~Ikeshoji, and C.~C. Liew, ``First-principles molecular-dynamics simulations of a hydrated electron in normal and supercritical water,'' {\em Physical review letters}, vol.~90, no.~22, p.~226403, 2003.

\bibitem{larsen2010does}
R.~E. Larsen, W.~J. Glover, and B.~J. Schwartz, ``Does the hydrated electron occupy a cavity?,'' {\em Science}, vol.~329, no.~5987, pp.~65--69, 2010.

\bibitem{uhlig2012unraveling}
F.~Uhlig, O.~Marsalek, and P.~Jungwirth, ``Unraveling the complex nature of the hydrated electron,'' {\em The Journal of Physical Chemistry Letters}, vol.~3, no.~20, pp.~3071--3075, 2012.

\bibitem{savolainen2014direct}
J.~Savolainen, F.~Uhlig, S.~Ahmed, P.~Hamm, and P.~Jungwirth, ``Direct observation of the collapse of the delocalized excess electron in water,'' {\em Nature chemistry}, vol.~6, no.~8, pp.~697--701, 2014.

\bibitem{uhlig2014optical}
F.~Uhlig, J.~M. Herbert, M.~P. Coons, and P.~Jungwirth, ``Optical spectroscopy of the bulk and interfacial hydrated electron from ab initio calculations,'' {\em The Journal of Physical Chemistry A}, vol.~118, no.~35, pp.~7507--7515, 2014.

\bibitem{alizadeh2015biomolecular}
E.~Alizadeh, T.~M. Orlando, and L.~Sanche, ``Biomolecular damage induced by ionizing radiation: the direct and indirect effects of low-energy electrons on dna,'' {\em Annual review of physical chemistry}, vol.~66, pp.~379--398, 2015.

\bibitem{herbert2019structure}
J.~M. Herbert, ``Structure of the aqueous electron,'' {\em Physical Chemistry Chemical Physics}, vol.~21, no.~37, pp.~20538--20565, 2019.

\bibitem{svoboda2020real}
V.~Svoboda, R.~Michiels, A.~C. LaForge, F.~Stienkemeier, P.~Slav{\'\i}{\v{c}}ek, and H.~J. W{\"o}rner, ``Real-time observation of water radiolysis and hydrated electron formation induced by extreme-ultraviolet pulses,'' {\em Science Advances}, vol.~6, no.~3, p.~eaaz0385, 2020.

\bibitem{kuhne2020cp2k}
T.~D. K{\"u}hne, M.~Iannuzzi, M.~Del~Ben, V.~V. Rybkin, P.~Seewald, F.~Stein, T.~Laino, R.~Z. Khaliullin, O.~Sch{\"u}tt, F.~Schiffmann, {\em et~al.}, ``Cp2k: An electronic structure and molecular dynamics software package-quickstep: Efficient and accurate electronic structure calculations,'' {\em The Journal of Chemical Physics}, vol.~152, no.~19, 2020.

\bibitem{pizzochero2019picture}
M.~Pizzochero, F.~Ambrosio, and A.~Pasquarello, ``Picture of the wet electron: a localized transient state in liquid water,'' {\em Chemical science}, vol.~10, no.~31, pp.~7442--7448, 2019.

\bibitem{zhang2020dp}
Y.~Zhang, H.~Wang, W.~Chen, J.~Zeng, L.~Zhang, H.~Wang, and E.~Weinan, ``Dp-gen: A concurrent learning platform for the generation of reliable deep learning based potential energy models,'' {\em Computer Physics Communications}, vol.~253, p.~107206, 2020.

\end{thebibliography}
